\documentstyle[epsfig]{article}
\begin{document}
\def\vphi{\langle \phi \rangle}
\def\tphi{\tilde{\phi}}
\def\bg{\beta{(g_s)}}
\def\s{\hat {s}}
\def\mt{m_t}
\def\mp{m_{\tphi}}
\def\mh{m_h}
\def\al{\alpha_s}
\def\d{\delta}
\def\sig{\sigma}
\def\mphi{m_{\phi}}
\def\gm{\gamma}

\begin{center}
{\Large \bf Production and Detection of Randall-Sundrum Light Radions
at Hadron Colliders}

\vskip 15pt

{\sf Uma Mahanta$^\dagger$ and Anindya Datta$^\heartsuit$} \\
{ Mehta Research Institute, Chhatnag Road,} \\
{Jhusi, Allahabad-211019, India}
\vskip .5cm
e-mail:  $^\dagger$ mahanta@mri.ernet.in, $^\heartsuit$
anindya@mri.ernet.in

\vskip .4 true in
{\bf Abstract}
\end{center}

In this paper we use the conformal anomaly in QCD to derive the coupling 
of light radion to gluons in the Randall-Sundrum model and use it to
compute the radion production cross section at hadron colliders by
gluon fusion.  We find that the radion production cross section by
gluon fusion at LHC would exceed that of the higgs boson by a factor
that lies between 7 and 8 over most of the range.  The decay modes of
the radion are similar to that of the SM higgs boson.  But the
striking feature is the enhancement of radion to 2-photon and radion
to 2-gluon branching ratio over the SM case. Utilising this, we then
discuss the possible search strategies of such scalars at Tevatron and
LHC.  Using the $\gamma \gamma$ decay mode one can explore/exclude
radion mass upto 1 TeV. Even with the current collected data at the
Tevatron, one can exclude radion mass upto 120 GeV for $\vphi$= 1 TeV.
\vskip 20pt
PACS number(s): 
\vskip 20pt

\vskip .5 in
\section{Couplings and Decay of the Radion to the SM particles}
Recently Goldberger and Wise showed that the modulus in the
Randall-Sundrum \cite{randal} scenario can be stabilized \cite{gold}
by introducing a scalar field in the bulk. The
stabilized modulus turns out to be very light if its mass arises from
a small bulk scalar mass. It was shown subsequently \cite{saki} that
the radion couples to the SM fields on the visible brane via the trace
of the energy momentum tensor.  In this article we shall study the
production and decay of such a light radion at hadron colliders.

The radion couples to the SM particles on the visible brane as given in 
\cite{saki}  {\it via} the relation,  ${\cal L}_{int}={1\over \vphi }
T^{\mu}_{\mu}\tphi$ where $T^{\mu}_{\mu}$ is the trace of the
symmetrized and conserved energy momentum tensor for SM fields. At the
tree level it is given by, 
\begin{equation}
T^{\mu}_{\mu} =\sum_f m_f\bar {f}f+2M^2_zZ^{\mu}Z_{\mu}+
2 M^2_w W^{\mu}W_{\mu}+2m_h^2 h^2.
\end{equation}
The fermion and gauge boson terms show the scale breaking effects due
to electroweak symmetry breaking.  Since the gluons and photons are
massless the radion does not couple to these at the tree level.
However the running of the gauge coupling in QCD and QED breaks the
scale invariance and induces a trace anomaly
\cite{collins}.  The trace anomaly in QCD therefore generates the
radion coupling to gluons which is given by

\begin{equation}
{\cal L}_{\tphi gg}={1\over \vphi }{\beta (g_s)\over 2 g_s}\tphi G^{a\mu\nu}
G^a_{\mu\nu}.
\label{rad-sm}
\end{equation}

where ${\bg\over 2g_s} =-(11-{2\over 3}n_f){g_s^2\over 32\pi^2}$.
$G^a_{\mu\nu}=[\partial_{\mu}g^a_{\nu}-\partial_{\nu}g^a_{\mu}
+gf^{abc} g^b_{\mu}g^c_{\nu}]$ is the gluon field strength tensor.
For $\mp ^2 <4 \mt^2 $ we have $n_f=5$ dynamical quarks and hence
${\bg\over 2g_s}\approx -3.84 {\alpha_s\over 4\pi}$. On the other hand
for $\mp ^2 > 4 \mt ^2 $ we have $n_f=6$ and hence ${\bg \over g_s}
\approx -3.50 {\alpha_s\over 4\pi}$.
The anomaly contribution is independent of fermion mass.
Even if EW symmetry were exact and all fermions had remained
massless the trace anomaly would still lead to the above coupling
of $\tphi$ to gluons. In the presence of EWSB, heavy quark loops give 
rise to a contribution to ${\cal L}_{\tphi gg}$. In the infinite
mass limit this contribution can be obtained by replacing $v$ by
$\vphi$ in the  effective Lagrangian for $hgg$ coupling [eqn. 10].
 It can be shown that
this contribution is smaller than the anomaly contribution written above.
The running of QED coupling also introduces a conformal anomaly. This gives 
rise to the following coupling of the radion to the photons.
\begin{equation}
{\cal L} _{\tphi \gamma \gamma} = \frac{1}{\vphi} \frac{\beta (e)}{2 e}
\tphi F _{\mu \nu} F ^{\mu \nu}
\end{equation}

Where $F ^{\mu \nu}$ has the usual meaning. We have used the following values 
of $\frac{\beta (e)}{2 e}$ in our calculation.
\begin{eqnarray}
\frac{\beta (e)}{2 e} &=& \frac{13 \alpha}{12 \pi} ~~~~~~~m_{\tphi} > 2 m _t 
\nonumber \\
&=& \frac{31 \alpha}{36 \pi} ~~~~~~~2 m _t > m_{\tphi} > 2 m _W \nonumber \\
&=&  \frac{10 \alpha}{9 \pi} ~~~~~~~m_{\tphi} < 2 m _W
\end{eqnarray}

We want to mention that in the SM, in the heavy quark limit, the $h 
 \gamma\gamma$ coupling is given by,

\begin{equation}
{\cal L} _{h \gamma \gamma} = \frac{1}{v} \frac{\beta (e)}
{2e} h F _{\mu \nu} F ^{\mu \nu}
\end{equation}

where,

\begin{equation}
\frac{\beta (e)}{2e} = \frac{2\alpha}{9 \pi} \nonumber
\end{equation}
Hence for $\mp ^2 , \mh ^2 < 4\;m_w^2$ (where $\tphi \rightarrow \gm
\gm$ is significant and the heavy top limit is valid) we have
${g_{\tphi \gm\gm}\over g_{h\gm\gm}}=5{v\over \vphi}$.  Although
$\tphi f\bar{f}$ and $\tphi VV$ ($V=W, Z$) couplings are suppressed
relative to $hf\bar{f}$ and $hVV$ couplings for $\vphi =$ 1 TeV,
$\tphi gg$ and $\tphi \gm\gm$ couplings are slightly enhanced relative
to $hgg$ and $h \gm\gm$ couplings.

\begin{figure}[htb]
\vspace*{-1.2in}
\centerline{\epsfig{file=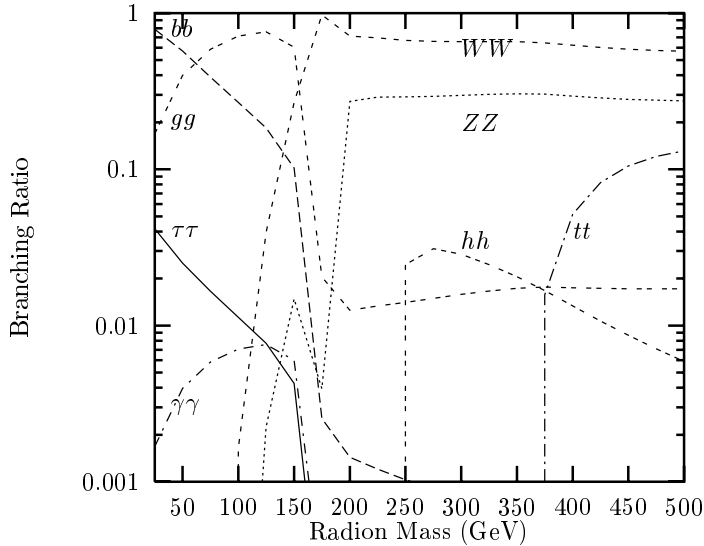,width=20cm}}
\vspace*{-5.9in}

\caption{\em Branching ratio of Radion into different channels for $\vphi 
=$ 1 TeV}
\end{figure}
	\label{fig_br}


Now we are in a position to present the radion decay branching ratios
into different channels. In fig. \ref{fig_br}, we present the relevant
decay branching ratios. For the illustrative purpose we assumed $\vphi
=$ 1 TeV. The radion coupling to any SM field has the 
$\frac{1}{\vphi}$ dependence. Thus if we choose a different value of
$\vphi$, the partial decay widths of different channels change in the
same fashion. But  the branching ratios remain
unchanged.  The striking
feature is that, for some radion masses the gluon gluon branching
ratio is almost equal to 1. The other thing to note is that the photon
photon branching ratio is also 5 to 8 times larger than the 2 photon
branching ratio of the SM higgs. We see in the next section that this
will have some interesting consequence in radion search. In this plot
the SM higgs mass of 120 GeV is used for the purpose of
illustration. This value of the SM higgs is well above the current
experimental lower bound from the LEP \cite{LEP_higgs}. 
Another interesting feature of this plot is that the radion branching
ratio into gg or $\gm\gm$ increases slowly with $\mp$ after the
sharp fall around WW threshold. On the contrary the higgs branching ratio
in these two modes decreases with $m_h$ after the WW threshold.

\section{Radion Production at Hadron Colliders}
The radion coupling to weak gauge bosons are suppressed relative
to $hWW$ and $hZZ$ couplings by a factor of ${v \over \vphi}$.
It is known that
higgs boson production at LHC by the weak boson fusion mechanism is
itself suppressed relative to the gluon fusion mechanism over most of
the range of $M_h$ ($M_h<$ 1 TeV). Hence the radion production at LHC
by weak boson fusion is also not expected to be the dominant or
efficient mechanism.  Also the radion couplings to light valence quarks
are extremely small. Therefore 
in this article we shall focus on radion production by gluon 
fusion.
The radion production cross-section at a hadron collider {\it via}
gluon fusion mechanism is given by,

\begin{equation}
\sig (p\;p({\bar p}) \rightarrow \tphi + X )= \Gamma (\tphi \rightarrow gg)\; 
\frac{\pi ^2}{8 \mp ^3}\; \tau 
\int_{\tau}^1 {dx\over x}\; g(x)\; g({\tau \over x})
\end{equation}

 Here $\tau$ is a dimensionless variable given by $\frac{\mp ^2}{S}$,
where $S$ is the proton - proton (anti-proton)   
center of mass energy. 
Note that $L_{\phi gg}$ generates momentum dependent $\tphi gg$,
$\tphi ggg$ and $\tphi ggg$ couplings. The strength of these couplings
are proportional to $\bg$ which includes the contribution of gluons as
well as the dynamical quarks. The Lagrangian also indicates that
radions could be produced either singly or in association with
gluons at a hadron collider by gluon fusion. In fact
 the gluon fusion 
mechanism turns out to be the dominant production process for radions
at hadron colliders over most of the interesting range.
 $qg\rightarrow \tphi$ and $\bar {q}g\rightarrow
\tphi$ does make a small contribution to radion production in 
$O(\alpha_s)^3$.  The radion coupling to gluons is very similar in
structure to the effective Lagrangian that gives the higgs coupling to
gluons [5] in the heavy quark limit. As in the case of the radion the
gluon fusion also turns out to be the primary production mechanism for
higgs boson at hadron collider. The dominant contribution to
$gg\rightarrow h$ arises from closed loops of heavy quarks that occur
in the theory. In this work we shall assume that the number of heavy
quarks ($N_h$) is equal to one namely the top.  It has been shown
\cite{sally} that the heavy quark limit ${M_h\over M_q}\rightarrow 0$
is an excellent approximation to the exact two loop corrected rate for
$gg\rightarrow h$.
 As $\mh\rightarrow 2\;\mt$ the exact result
rises above the heavy quark result and exhibits a small bump
corresponding to the $t\bar {t}$ threshold. The width of the bump {\it
i.e} the departure region increases with increasing $\mt$. The
disagreement between the two results in the $\mh \ge 2\mt$ region
however is always less than a factor of two at LHC. The heavy quark
limit for $gg \rightarrow h$ can be obtained from the gauge invariant
effective Lagrangian \cite{van}

\begin{equation}
{\cal L} = -\;{1\over 4}\;\bigg[ 1 - {2\;\beta_h \over g_s(1+\delta )}
\;{h\over v}\bigg]\; G^a_{\mu\nu} G^{a\mu\nu}-{\mt\over v}\;h\bar {t}t.
\end{equation}

 $\delta =1+2 {\alpha_s\over \pi}$ is the anomalous dimension of the
 mass operator arising from QCD interactions. $\beta_h$ is the heavy
 quark contribution to the QCD beta function. Since the $hgg$ coupling
 in the $M_q\rightarrow \infty$ limit arises from heavy quark loops it
 is only the heavy quarks that contribute to the $\beta_h$ in eqn(3).
 To order $(\al^3)$ the heavy quark contribution [6] to $\beta (g_s)$
 is given by $\beta_h=N_h{\al\over 12\pi}[1+{19\al\over 4}]$. On the
 other hand the $\bg $ that appears in the gluon coupling to the
 dilaton like radion mode arises from the trace anomaly. The trace
 anomaly has its origin in the heavy regulator fields as their masses
 are taken to infinity.  So it includes the gluonic contribution as
 well as that of dynamical quarks.  This difference in the two beta
 function contributions makes the $\tphi gg$ coupling greater than the
 $hgg$ coupling even for ${v\over \vphi}={1\over 4}$.  However with
 increasing $\vphi$ the $hgg$ coupling ultimately wins over the $\tphi
 gg$ coupling. This feature is clear from fig. \ref{lhc} where it is
 shown that with increasing $\vphi$, $\sigma (pp\rightarrow \tphi )$
 ultimately becomes smaller than $\sigma (pp\rightarrow h )$.

Let us now make some rough numerical estimate about the ratio 
${\sigma (pp\rightarrow \phi )\over \sigma (pp\rightarrow h )}$ in the
lowest order (O($\al $)). Our estimates will depend only on the
relative strength of $\phi gg$ and $hgg$ couplings. For $\sqrt {\s}<
2\mt$, ${\bg \over 2g_s}=-3.84{\al\over 4\pi}$ whereas $\beta_h={1\over 3}
{\al\over 4\pi}$ to lowest order. Also in this region the heavy quark
limit provides a good approximation to the exact result for 
$\sig (pp\rightarrow h)$. We find that ${g_{gg\phi}\over g_{ggh}}
\approx -2.88$ for $\vphi =$ 1 TeV. Above the $2\mt$ threshold
we have ${\bg\over 2g_s}\approx -3.50{\al\over 4\pi}$. Although in this
region the heavy quark limit does not work that well for the higgs cross
section we can still get an order of magnitude estimate 
(lower by at most a factor of two) using it. The ratio of couplings
 now ($\sqrt {\s}> 2\mt$) becomes ${g_{gg\phi}\over g_{ggh}}
\approx -2.63$. 
\begin{figure}[htb]
\vspace*{-1.2in}
\centerline{\epsfig{file=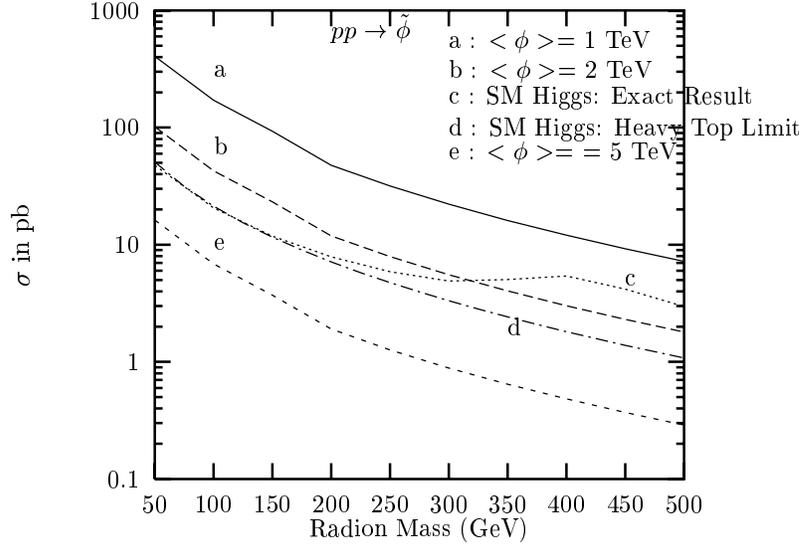,width=20cm}}
\vspace*{-5.9in}

\caption{\em Radion production cross-section (a,b,e) at the LHC, for 
different values of $\vphi$. We also present the SM higgs production 
cross-section (c) the exact result and (d) in the heavy quark limit.}
\end{figure}
	\label{lhc}


Using the fact that the effective Lagrangian for higgs and
radion production by gluon fusion are similar except for couplings we
find that if $\vphi $=1 TeV then the cross section for $pp\rightarrow
\tphi$ will exceed that of $pp\rightarrow h$ by a factor of 8.3 for
$\sqrt {\s}<2\mt$ and by a factor of 6.9 for $\sqrt {\s}>2\mt$.
However for $\vphi $=5 TeV the radion production cross section will be
suppressed relative to the higgs cross section roughly by a factor of
three. These features have been exhibited in fig. \ref{lhc} where we
have plotted the lowest order higgs production cross section (both
exact and the heavy quark limit) and the radion cross section (for
three different values of $\vphi$) against the mass of the particle
($\tphi$ or h) at LHC.

The estimates given above are based on lowest order calculations. 
It is known that
higher order QCD corrections increases the lowest order rate by a
factor (K factor) that lies between 2 and 3 at LHC \cite{sally,abdel}.
The QCD radiative corrections done in the heavy quark limit forms an
excellent approximation to the exact calculations. To calculate the K
factor one therefore always uses the heavy quark limit.  But in the
heavy quark limit the effective Lagrangian for higgs production is
similar in structure to the Lagrangian for radion production.  Hence
the K factor for higgs production in the heavy quark limit will be the
same for the radion also. So higher order QCD corrections will not
affect the relative rate between the radion and the higgs to a very
high degree of accuracy.

\section{Detection and Possible SM Backgrounds}

Let us now concentrate on the detection of radion at a hadron collider
like Tevatron.  The dominant decay mode of a 50-150 GeV radion as can
be seen from fig. \ref{fig_br}, is to $b {\bar b}$ or to $gg$. But the
 striking feature is that the $\gamma \gamma$ branching ratio of
the radion is larger than the higgs case by a factor 5-8, over a
considerable mass range. The higgs production rate by gluon fusion 
and and its decay into the
$\gm\gm$ mode are both suppressed relative to that of the radion.
This is the reason why two photon final state is not a good bet for
the higgs at the Tevatron.

At the Tevatron radion production cross-section varies from 140 $pb$
to 1 $pb$ as we vary radion mass from 20 (current lower bound on radion
mass comes from LEP -II \cite{uma_subhe}) to 160 GeV. We have included
a NLO QCD correction factor of 2 in our calculation. This
cross-section with the presently collected luminosity will give rise to some
10 $^4$ radions. If the radion decays into the $\gm\gm$ mode the final
state will consist of 2 hard photons. The main background of this
2-photon final state comes from the pair annihilation of the valence
quarks and anti-quarks.  The other dominant source of $\gamma \gamma$
background is gluon gluon annihilation to two photons. Though this is
suppressed to the former by a factor of $\alpha _s ^2$, dominance of
gluon flux over the quarks flux, can make this comparable with the
former.  We do not calculate this second contribution explicitly. We
multiply the $q {\bar q} \rightarrow
\gamma \gamma$ contribution by a factor of 2 to take this into
account. At the Tevatron this is a conservative approximation.  We
have used a parton level monte-carlo event generator to estimate the
numbers for both the signal and the background  and the CTEQ-4M
\cite{CTEQ} parametrisation for the parton densities in
our entire analysis.

The following cuts have been applied to differentiate between signal and 
background.

\noindent
$
p_T^{\gamma} > 10$ GeV

\noindent
We demand the photons  are in the central part of the detector.

\noindent
$|\eta_{\gm}| < 3.$.

\noindent
We also require that the angular separation between the photons be
substantial, {\em i.e.}
\noindent
$ \Delta R_{jj}
\equiv \sqrt{ (\Delta \eta_{\gamma \gamma })^2 + 
(\Delta \phi_{\gamma \gamma})^2 } > 0.5
$

\begin{figure}[htb]
\vspace*{-1.2in}
\hspace*{-0.5in}
\centerline{\epsfig{file=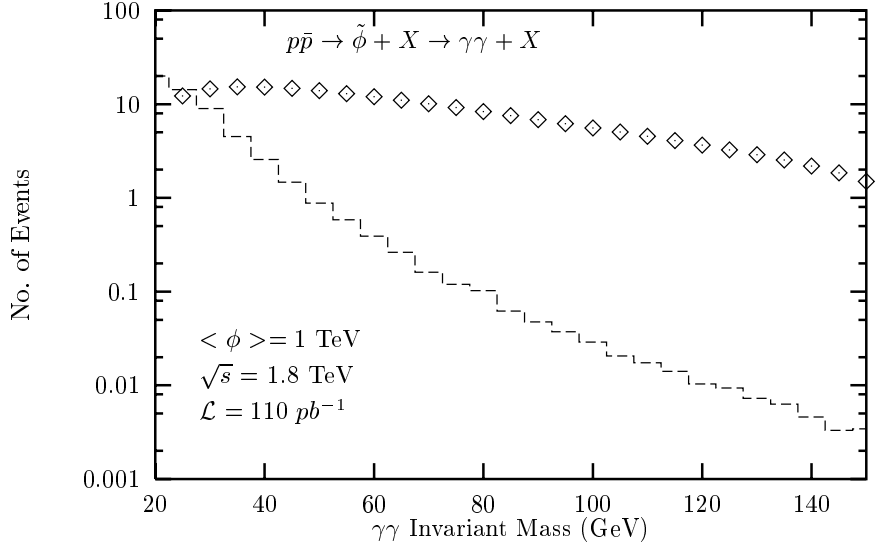,width=20cm}}
\vspace*{-5.9in}

\caption{\em Invariant mass distribution for signal and background at the
Tevatron for $\vphi =$ 1 TeV.}
\end{figure}
	\label{tev_100pb}

Even  after applying these cuts, the SM background cannot be removed
completely.  So the strategy is to compare the invariant
mass (of the photon pair) distribution of the signal and background.
For the signal (of a definite $\mp$), invariant mass distribution
shows a sharp peak over the continuum background. The sharpness of the
peak depends mainly on the detector resolution and the partial width of
$\tphi \rightarrow \gamma \gamma$. In fig. \ref{tev_100pb} we
plot the invariant mass ($M _{\gamma \gamma}$) distribution for the
signal (dots) and backgrounds (broken histogram) assuming an uniform
bin size of 5 GeV.  Here one can see that, for low $M _{\gamma
\gamma}$, number of background events is higher than the signal. But
as $M _{\gamma \gamma}$ increases,  the number of background events in the
mass bins falls off more rapidly than the signal- which more or
less remains the same over the entire mass range we are interested
here. Once the radion mass is near to $2 m_w$, signal events falls off
sharply, due to the sharp fall of $\gamma \gamma$ branching ratio.

 We find that the $\gm \gm$ mode is good enough to exclude radion mass
 nearly upto 120 GeV even with the presently collected luminosity of
 110 $pb ^{-1}$. Here we have taken the radion $vev$ to be 1 TeV.  On
 the other hand at the Tevatron with the presently collected
 luminosity one cannot say anything about the SM higgs.

Now let us go over to the case of Upgraded Tevatron with center of
mass energy of 2 TeV and luminosity of 1 $fb ^{-1}$. As center of mass
energy is almost equal to the present, the signal and background
 are just 10 times larger than the previous case. This is
evident from the fig. 4. In this figure we also plot the
significance ($\equiv \frac{No. of Signal}{\sqrt{No. of Background}})$
of our signal upto $M _{\gamma \gamma}$ equal to 100 GeV. One can check, 
over this entire mass range, significance is greater than 5. Generally
a significance greater than 5 points to the discovery. One can also
easily check that once $M _{\gamma \gamma}$ is greater than 100 GeV,
no. of background events in the corresponding bins become less than 1
events. So in this region we can demand 5 signal events as a benchmark
for discovery. Thus at the upgraded Tevatron one can discover radion mass
upto 160 GeV and exclude upto 165 GeV. Once $M _{\gamma \gamma}$ is
greater than 165 GeV, our signal falls off very sharply and we cannot
say anything more about it.

\begin{figure}[htb]
\vspace*{-1.2in}
\hspace*{-0.5in}
\centerline{\epsfig{file=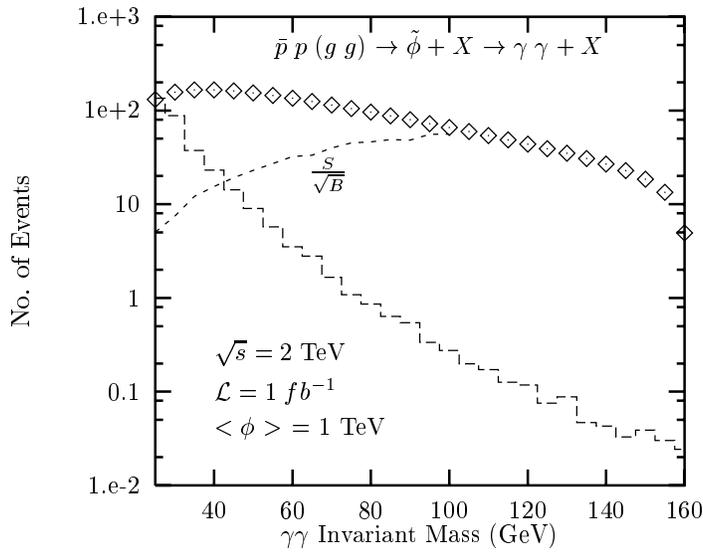,width=20cm}}
\vspace*{-5.9in}

\caption{\em Invariant mass distribution for signal and background at the
Tevatron Upgrade for $\vphi =$ 1 TeV.}
\end{figure}
	\label{fbtev}


 If we change $\vphi$ the branching ratio of the radion to different
 channels remains same.  So if we take $\vphi =$ 2 TeV, the
 cross-section and number of $\gm \gm$ events become $\frac{1}{4}$ of
 the present case ($\vphi =$ 1 TeV). And at the Tevatron with the
 presently collected luminosity, we cannot say anything about it. At
 the Tevatron Upgrade, we also cannot talk about the discovery, for
 $\mp < $45 GeV. At higher masses, one can discover upto 150 GeV
 $\mphi$ if $\vphi =$ 2 TeV instead of 1 TeV.

Finally we want to examine the search prospects of this scalar
particle at the LHC. 
 Though the two photon
branching ratio drops very sharply near $\mp \sim$ 140 GeV, in
contrast to the SM case it remains constant with $\mp$ after 140 GeV. 
Also the radion production rate is almost 8 times higher than that of
higgs production rate. So unlike the SM higgs boson, for $\mp >$ 140
GeV, $\gamma \gamma$ mode is a viable avenue
to discover or exclude the radion for
masses well upto 1 TeV. 

\begin{figure}[htb]
\vspace*{-1.2in}
\hspace*{-0.5in}
\centerline{\epsfig{file=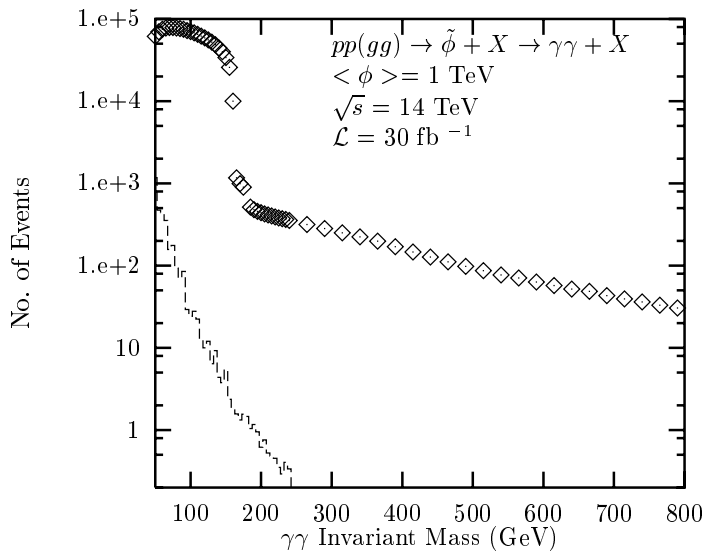,width=20cm}}
\vspace*{-5.9in}
\caption{\em Invariant mass distribution for signal and background at the
LHC for $\vphi =$ 1 TeV.}
\end{figure}
	\label{lhc_evnts}

Conservatively, we take the luminosity to be
equal to 30 $fb ^{-1}$. The source of SM backgrounds remain the same
here. But unlike the Tevatron, the 2-photon background coming from pair
annihilation of quark and anti-quark becomes less severe. This is because
 the anti-quark
coming from one proton has to be excited from the sea. But gluon density in
the proton is much larger than the quark density at LHC. So to take
into account the gluon gluon contribution to 2-photon background we
multiply the quark-anti-quark contribution by a factor of 6 \footnote{
Generally at the LHC energies the gluon-gluon contribution to two
photon background is 5 times larger than the quark anti-quark
contribution. For $m_{\gamma\gamma} = $ 110 (130) GeV this factor has
been estimated to be 3.55 (3.56) for the LHC \cite{CMS}.}. The cuts we
used here are rather similar to the Tevatron case. Only we change the
cuts on the transverse momenta of the photons to
 be greater than 20 GeV. We
present in fig. 5 the number of signal events and the
corresponding number for the background against the $\gamma \gamma$
invariant mass. To plot the invariant mass distribution for the
background we choose a bin size of 5 GeV. Form the figure one can
easily see that the background is an order of magnitude smaller than the
signal over the entire mass range. And once $M _{\gamma
\gamma}$ is greater than 250 GeV, there are less than one event in the 
respective mass bins. That's why we do not show it in the figure. So
LHC can easily discover radions with masses upto 1 TeV for $\vphi$=1
Tev.  If we increase $\vphi$ to 4 TeV one can easily check from fig. 5
that our discovery limit comes down to 650 GeV.

\section{Conclusions}
In this paper we examined the radion production and its subsequent
decay into SM particles at hadron colliders. The radion production
cross-section is larger than the higgs production cross-section by
a factor of 6-8.  The partial width of the
radion to two gluons and two photons is also
enhanced due to the enhanced $\tphi\;g\;g$ or $\tphi\;\gamma\;\gamma$
coupling. The other partial widths are suppressed with respect
to  higgs widths by a factor that depends on $\vphi$. We also
discussed the viability of the 2-photon signal for the radion at the
Tevatron. One can exclude radion mass upto 120 GeV even with the
presently collected luminosity. Upgraded Tevatron with higher center
of mass energy and higher luminosity certainly can discover radion
upto 160 GeV mass, if not it can exclude it upto 165 GeV. 
Similarly at the LHC one can definitely discover radions with
masses up to 1 
TeV. These estimates are done assuming $\vphi$ =1 TeV. With increasing
$\vphi$ both the exclusion limit and the discovery limit using
the $\gm\gm$ mode will come down. 

\vskip .5in
{\bf Note added: While this work was being completed there appeared 
one paper \cite{giud} which discusses some of the issues presented here.}

\end{document}